%% file: main.tex
\documentclass{article}

\usepackage{PRIMEarxiv}

\usepackage[utf8]{inputenc} 
\usepackage[T1]{fontenc}    
\usepackage{hyperref}       
\usepackage{url}            
\usepackage{booktabs}       
\usepackage{amsfonts}       
\usepackage{nicefrac}       
\usepackage{microtype}      
\usepackage{lipsum}
\usepackage{fancyhdr}       
\usepackage{graphicx}       
\graphicspath{{media/}}     

\usepackage{graphicx}
\usepackage{wrapfig}
\usepackage{algorithm, algpseudocode}
\usepackage{amssymb}
\usepackage{booktabs}
\usepackage{makecell}
\usepackage{tikz}
\usepackage[table]{xcolor}
\usepackage{markdown}

\usepackage{listings}
\usepackage{subcaption}
\definecolor{lightgreen}{RGB}{242,250,242}
\definecolor{listingbg}{RGB}{248,248,248}
\definecolor{listingframe}{RGB}{210,210,210}
\lstdefinestyle{appendixbox}{
  basicstyle=\ttfamily\scriptsize,
  breaklines=true,
  breakatwhitespace=false,
  columns=fullflexible,
  keepspaces=true,
  showstringspaces=false,
  frame=single,
  framerule=0.35pt,
  rulecolor=\color{listingframe},
  backgroundcolor=\color{listingbg},
  xleftmargin=0.5em,
  xrightmargin=0.5em,
  aboveskip=0.75em,
  belowskip=0.75em,
  literate={—}{{---}}1 {–}{{--}}1 {'}{{'}}1 {"}{{``}}1 {"}{{''}}1 {tau}{{$\tau$}}1 {^2}{{$^2$}}1 {bullet}{{$\bullet$}}1
}

\newcommand{\ie}{\textit{i.e.}}
\newcommand{\eg}{\textit{e.g.}}

\definecolor{mygreen}{HTML}{94e66c}

\usepackage[dvipsnames]{xcolor}
\newcommand{\low}[1]{\cellcolor{mygreen!10}{#1}}
\newcommand{\med}[1]{\cellcolor{mygreen!25}{#1}}
\newcommand{\high}[1]{\cellcolor{mygreen!40}{#1}}
\newcommand{\vhigh}[1]{\cellcolor{mygreen!65}{#1}}

\usepackage[table]{xcolor}
\usepackage{booktabs}
\usepackage{array}
\usepackage{fontawesome5}

\usepackage[most]{tcolorbox}
\tcbuselibrary{listings,breakable,skins}
\usepackage{xcolor}

\newcommand{\squishlist}{
 \begin{list}{$\bullet$}
 		{ \setlength{\itemsep}{0pt}
 			\setlength{\parsep}{3pt}
 			\setlength{\topsep}{3pt}
 			\setlength{\partopsep}{0pt}
 			\setlength{\leftmargin}{1.5em}
 			\setlength{\labelwidth}{1em}
 			\setlength{\labelsep}{0.5em} } }
\newcommand{\squishend}{
  \end{list}  }

\usepackage{tcolorbox}
\tcbuselibrary{skins} 

\newcommand{\codeblk}[1]{%
  \begingroup
  \setlength{\fboxsep}{2pt}
  \colorbox{gray!20}{\small\ttfamily\detokenize{#1}}%
  \endgroup
}

\newtcolorbox{promptbox}[1][]{
  breakable,
  enhanced,
  colback=blue!3,
  colframe=blue!40!black,
  boxrule=0.4pt,
  arc=4pt,
  left=8pt,
  right=8pt,
  top=8pt,
  bottom=8pt,
  width=\columnwidth,
  fontupper=\normalsize,
  coltitle=white,
  colbacktitle=blue!50!black,
  fonttitle=\bfseries,
  title=#1
}

\title{Thinking Ahead: Prospection-Guided Retrieval of Memory \\ with Language Models}

\author{
  Harshita Chopra\textsuperscript{1} \quad
  Krishna Chintalapudi\textsuperscript{2} \quad
  Suman Nath\textsuperscript{2} \quad
  Ryen White\textsuperscript{2} \quad
  Chirag Shah\textsuperscript{1} \\ [10pt]
  \textsuperscript{1}University of Washington, Seattle, USA \\
  \textsuperscript{2}Microsoft Research, Redmond, USA \\ [10pt]
  \small Email: \texttt{hchopra@cs.washington.edu} \quad \quad \href{https://github.com/harshita-chopra/PGR-mem}{\faGithub\ github.com/harshita-chopra/PGR-mem}
}

\begin{document}
\maketitle

\input{abstract}
\input{introduction}

\input{relatedwork}
\input{pgr}
\input{benchmark}

\input{evaluation}
\input{conclusion}

\bibliographystyle{unsrt}  
\bibliography{references}  

\newpage
\appendix
\section{Appendix}
\label{sec:appendix}
\input{appendix}

\end{document}

%% file: abstract.tex
\begin{abstract}
    Long-horizon personalization requires dialogue assistants to retrieve user-specific facts from extended interaction histories.
    In practice, many relevant facts often have low semanticsimilarity to the query under dense retrieval.
    Standard Retrieval-Augmented Generation (RAG) and GraphRAG systems are still largely retrospective: they rely on embedding similarity to the query or on fixed graph traversals, so they often miss facts that matter for the user's needs but lie far from the query in embedding space.
    Inspired by \emph{prospection}, the human ability to use imagined futures as cues for recall, we introduce \textbf{Prospection-Guided Retrieval (PGR)}, which decouples retrieval from how memories are stored.
    Given a user query, PGR first expands the goal into a short Tree-of-Thought (ToT) or linear chain of plausible next steps, and uses these steps as retrieval probes rather than relying on the original query alone.
    The facts retrieved by these probes are then used to personalize the next round of prospection, enabling PGR to uncover additional memories that become relevant only after the simulation is grounded in the user's history.
    We also introduce \textbf{MemoryQuest}, a challenging multi-session benchmark in which each query is annotated with $3$--$5$ dated reference facts subject to a low query--reference similarity constraint.
    Across $1{,}625$ queries spanning $185$ user profiles from 3 publicly available datasets, PGR-TOT substantially improves retrieval, including nearly 3x recall on MemoryQuest over the strongest baseline.
    In pairwise LLM-as-judge comparisons against baselines, PGR-generated responses are preferred on $89$--$98\%$ of queries, with blinded human annotations on held-out subsets showing the same trend.
    Overall, the results demonstrate that explicit prospection yields large gains in long-horizon retrieval and response quality relative to similarity-only baselines.
\end{abstract}

%% file: introduction.tex
\section{Introduction}
\label{sec:introduction}


Human memory is not merely a lookup table; it is a simulation engine. We do not retrieve memories solely by matching the present to the past. Rather, neuroscientific research has established that the brain employs prospection—the simulation of possible futures to trigger the recall of memories that will become relevant \cite{schacter2017episodic,addis2007constructive}. This mechanism allows us to anticipate obstacles and surface "latent" memories that seem irrelevant to the present moment but become critical for an anticipated future state. While human intelligence relies on a dynamic dialogue between the prefrontal cortex and the hippocampus to iteratively simulate and refine possible futures and retrieve memories, modern AI systems remain anchored to a lookup mentality, ignoring memory's predictive utility entirely.

\vspace{-0.2cm}
\paragraph{Decoupling Retrieval from Storage.} Current Retrieval-Augmented Generation (RAG) \cite{guu2020retrieval, lewis2020rag} and GraphRAG \cite{edge2024graphrag} systems suffer from a fundamental coupling of \textit{storage structure} and \textit{retrieval strategy}. In these paradigms, retrieval is a mere byproduct of the architecture: vector stores mandate similarity search, while graphs necessitate traversal. We argue that retrieval should be an independent policy. By decoupling how memories are stored from how they are retrieved, PGR enables agents to reason about informational needs in advance.

Inspired by the anticipatory mechanisms of the human brain, we introduce \textit{\textbf{Prospection-Guided Retrieval (PGR)}}, a novel retrieval paradigm for personalized agentic memory. Unlike existing methods that retrieve retrospectively by matching a query to past records, PGR retrieves \textit{prospectively}: given a user goal, it operationalizes the dialogue between foresight and recall through an iterative reasoning loop:
\begin{align}
    \text{Simulate Future} &\rightarrow \text{Retrieve Relevant Memories} \hspace{-9em} &\rightarrow \text{Refine Trajectory}
\end{align}
Crucially, PGR is a \textit{retrieval policy, not a storage mechanism}. It is orthogonal to the underlying memory representation and can be composed with vector stores, entity graphs, or episodic logs to transform static recall into anticipatory intelligence. Because simulation occurs in real-time, PGR avoids the limitations of the rigid, pre-indexed structures like GraphRAG, allowing the system to dynamically adapt and surface latent memories that a static search would fail to reach.

\begin{figure}[ht]
            \centering
            \centering
            \includegraphics[width=\columnwidth]{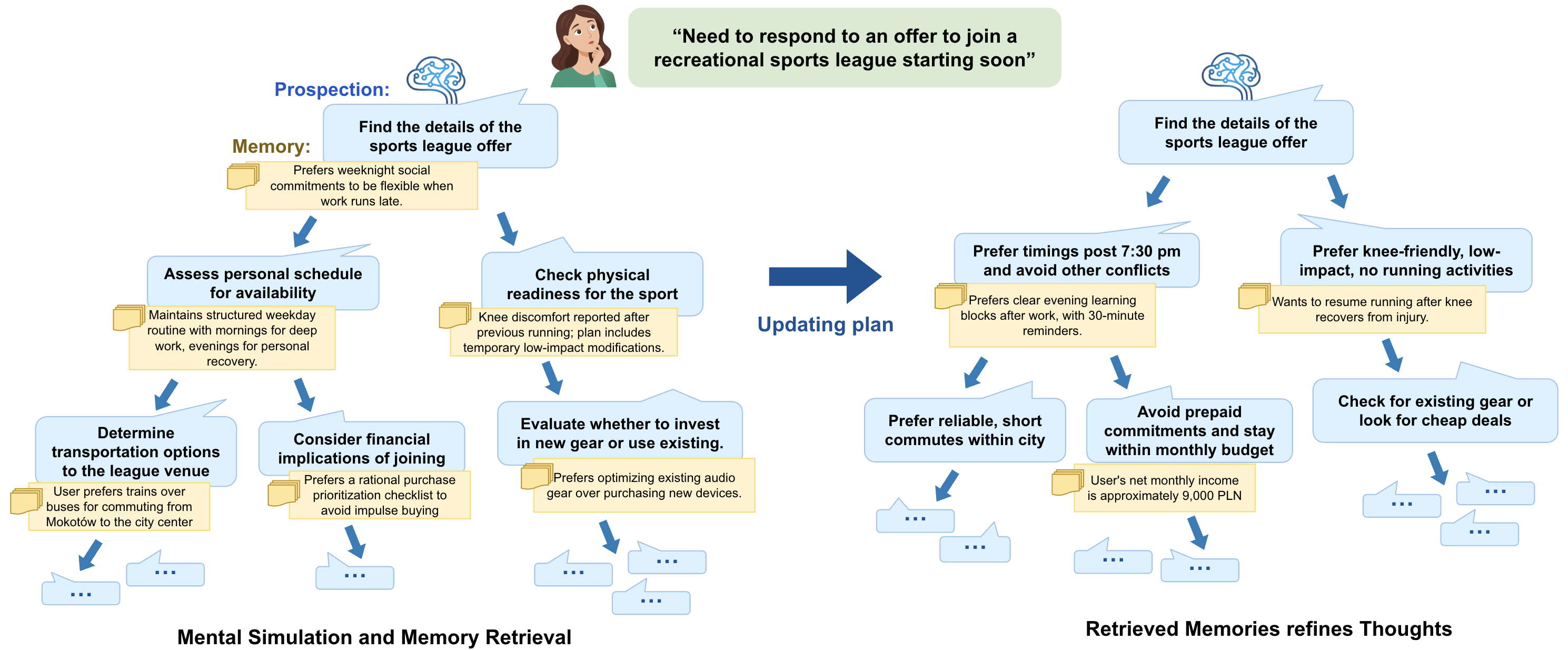}
            \caption{Prospection involves simulating possible futures and triggering recall of context-relevant past memories at each imagined step.  
    These forward simulations enable anticipatory and contextually appropriate behavior. Retrieved memories modify the simulated plan accordingly.}
            \label{fig:intro-prospection}
\end{figure}

Figure~\ref{fig:intro-prospection} illustrates a trace of the iterative loop in PGR from our dataset. When a user considers joining a recreational sports league, the system executes a \textbf{Tree-of-Thought (ToT)} to simulate parallel future trajectories, such as assessing schedule availability or physical readiness. Each hypothesized step triggers the recall of latent past experiences (e.g., knee discomfort from prior running or a preference for weeknight flexibility). These retrieved memories iteratively refine the ``mental simulation''; for instance, recalling a knee injury shifts the prospection from high-impact activities to low-impact alternatives. 

By entertaining multiple futures simultaneously, PGR predicts informational needs before they are encountered. This allows it to surface indirect or counterfactual memories, such as a specific transportation preference for a city center venue, that conventional similarity-based retrieval would miss because they share no surface-level overlap with the initial query.

To evaluate PGR, we introduce \textbf{MemoryQuest}, a novel benchmark designed for multi-session, multi-task human--agent interactions. Sessions are chronologically consistent, with events logically evolving over weeks to reflect realistic long-term dependencies. Unlike datasets such as \emph{MSC}~\cite{xu-etal-2022-beyond} or \emph{LongMemEval}~\cite{wu2024longmemevalbenchmarkingchatassistants}, which often focus on direct semantic matches, MemoryQuest enforces queries that require retrieval of multiple $3$--$5$ memories semantically dissimilar from the query scattered across multiple domains. MemoryQuest serves as a stress test for long-term personalization, exposing the failure modes of standard RAG that prospection-based methods are designed to overcome.

Our key contributions are:
\squishlist
\item \textbf{Prospection-Guided Retrieval (PGR):} A brain-inspired paradigm that decouples retrieval policy from storage via an iterative $\text{Simulate} \rightarrow \text{Retrieve} \rightarrow \text{Refine}$ loop. This enables agents to surface latent or counterfactual memories by anticipating future needs, overcoming the rigidity of pre-ordained structures like GraphRAG.
\item \textbf{MemoryQuest Benchmark:} A multi-session dataset designed for long-term personalization. It requires retrieving $3$--$5$ disparate memories linked by chronological and logical dependencies rather than surface semantic similarity, specifically targeting the "retrospective bottleneck" of current systems.
\item \textbf{Empirical Evaluation:} A scalable evaluation framework using human-as-a-judge (MTurk) to ground and validate a large-scale LLM-as-a-judge analysis across 1,500+ queries on three benchmarks. Results demonstrate that PGR significantly outperforms baselines in retrieval recall and anticipatory response quality.
\squishend

%% file: relatedwork.tex
\section{Related Work}
\label{sec:relatedWork}

\noindent
{\bf Retrieval-Augmented Generation (RAG).}
Standard RAG~\cite{lewis2020rag} integrates external knowledge into the generation process to improve factual grounding. While subsequent hybrid variants~\cite{izacard2021fid} and persistent memory stores~\cite{park2023generative} improve scalability, they remain inherently \emph{retrospective}. These systems retrieve based on surface similarity to the current query rather than goal-oriented anticipation, often failing to surface memories that are only indirectly relevant to future needs.

\noindent
{\bf Graph-Based Retrieval and GraphRAG.}
To move beyond isolated text chunks, \textit{GraphRAG}~\cite{edge2024graphrag} leverages knowledge graphs to support multi-hop inference. However, these methods are constrained by a \emph{preordained structure}: retrieval is limited to traversing static edges established at construction time. Consequently, they cannot infer counterfactual relations or anticipate connections dynamically based on specific query -- a hallmark of human prospection.

\noindent
{\bf Agentic Memory and Reasoners.}
Recent frameworks such as \textit{ToT}~\cite{yao2023tree} and \textbf{ReAct}~\cite{yao2023react} use explicit reasoning traces to enhance planning and action. Agentic memory systems such as Mem0$^{\texttt{g}}$~\cite{chhikara2025mem0} maintain persistent user memories and entity relations, but retrieval still relies on vector search and extracted graph structure. In contrast, \textit{PGR} extends the principles of branching reasoning to the \emph{retrieval domain}: simulated ToT or chain steps become retrieval probes, and retrieved facts refine subsequent prospection. This transforms retrieval into an active, iterative policy that is orthogonal to—and can be layered upon—any underlying storage architecture.

\noindent
{\bf Neuroscience of Prospection.}
Cognitive neuroscience observes that human memory is constructive, not archival; hippocampal-prefrontal circuits support both remembering the past and imagining the future~\cite{schacter2017episodic, addis2007constructive}. This \emph{simulation} allows fragments of experience to be recombined for anticipatory planning~\cite{hassabis2007scenes}. PGR operationalizes this insight, aligning memory retrieval with active simulation mechanisms observed in human cognition.

%% file: pgr.tex
\section{Prospection-Guided Retrieval}
\label{sec:method}

\begin{figure*}[t]
\centering
            \begin{minipage}[t]{0.9\textwidth}
            \vspace{-0.1in}
            \includegraphics[width=\columnwidth]{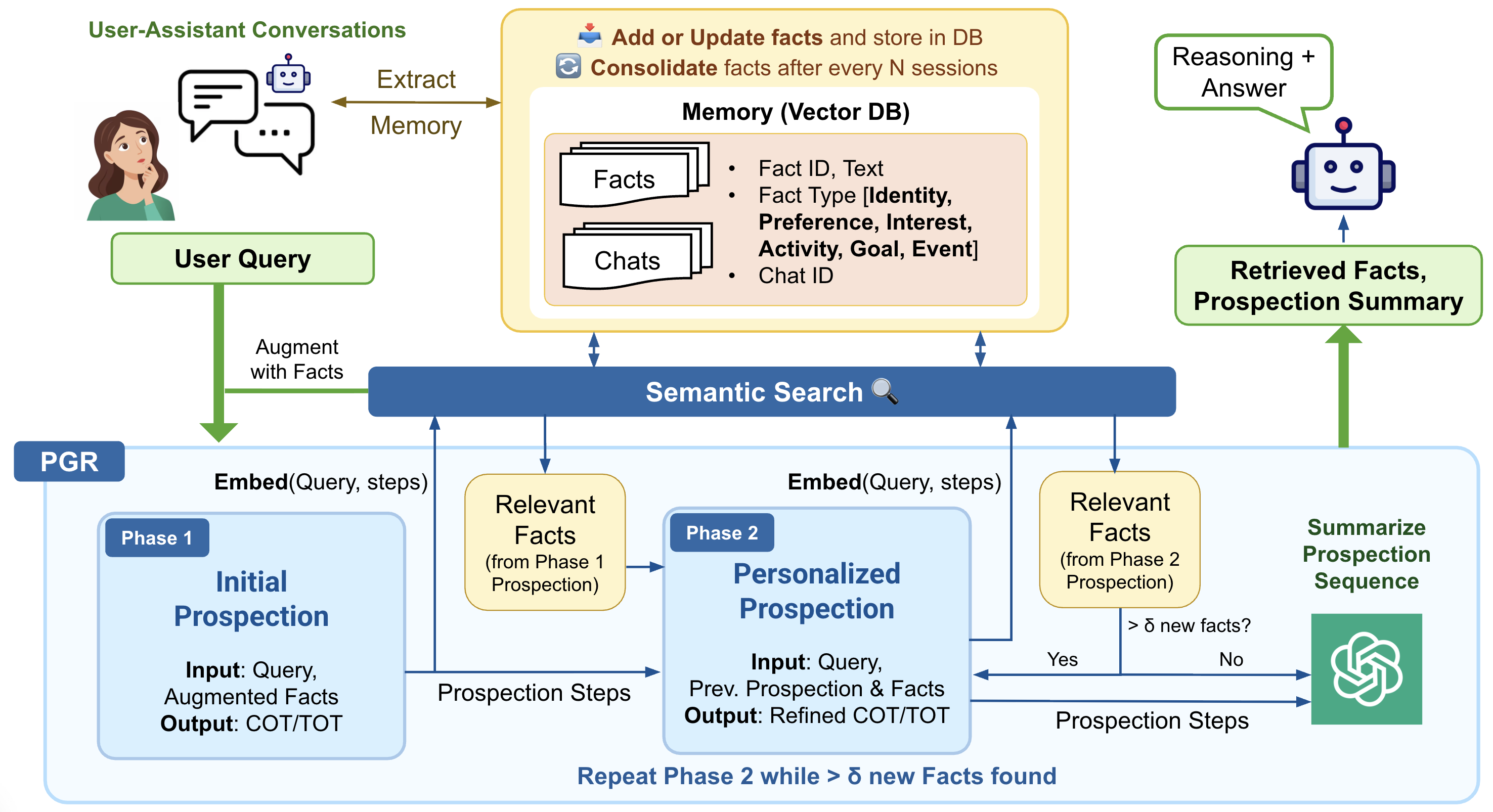}
            \vspace{-0.2in}
            \caption{\small{Overview of PGR}}
            \label{fig:pgr-fig}
            \end{minipage}
\end{figure*}

Inspired by the neurobiological dialogue between the prefrontal cortex and the hippocampus, Prospection-Guided Retrieval (PGR) operationalizes \emph{prospection}---the mental simulation of possible futures used to trigger the recall of contextually relevant memories~\cite{schacter2017episodic}. Unlike RAG and GraphRAG, which are anchored to a static ``lookup mentality'' dictated by storage structure, PGR treats memory as a substrate managed by an independent retrieval policy. Acting as an executive controller, PGR executes an iterative $\text{Simulate} \rightarrow \text{Retrieve} \rightarrow \text{Refine}$ loop to surface latent or counterfactual memories. This framework transforms archival records into a dynamic resource for real-time anticipatory intelligence.

\subsection{System Architecture Overview}
\label{sec:architecture}

As illustrated in Figure~\ref{fig:pgr-fig}, our architecture comprises three interconnected modules: the \textbf{Chat Interface}, \textbf{Long Term Memory}, and \textbf{PGR}.

\vspace{-0.2cm}
\paragraph{Chat Interface.} 
The assistant interacts with the user via a standard dialogue interface. Queries are sent to PGR, which returns relevant facts and a prospection reasoning summary. This summary enables the agent to generate responses that are both factually grounded and anticipatory.

\vspace{-0.2cm}
\paragraph{Long Term Memory.} 
Analogous to the hippocampus, this layer maintains long-term knowledge across sessions. Our reference implementation uses a structured fact-store with periodic consolidation, but the system is substrate-agnostic. PGR accesses it via a standard query interface (e.g., vector or graph similarity) as a black-box resource for simulations.

Serving as a computational prefrontal cortex, PGR mediates between interface and memory. It executes an iterative retrieval process, generating future trajectories and grounding them in memory to identify contextually vital information. The module returns both retrieved memories and the prospection reasoning that guided their selection, enabling the agent to produce anticipatory responses.

\subsection{Long Term Memory}
\label{sec:pgr-memory}
The memory store maintains the agent's knowledge of the user across sessions. PGR is agnostic to the underlying storage technology and requires only a queryable memory interface; the backend may be a vector index, a knowledge graph, or a hybrid store. Our implementation uses two structures: \emph{conversation logs} and \emph{memory facts}.

\vspace{-0.2cm}
\paragraph{Conversation logs.} Each user--agent session is stored as a multi-turn conversation with a unique id and date. These logs merely serve as the source for memory fact extraction but not used for retrieval.

\vspace{-0.2cm}
\paragraph{Memory facts.} Facts are short, atomic statements extracted from conversations \eg demographics, preferences, plans, past events. Each fact is associated with a unique id, a type, a frequency count, related entities, and provenance (the sessions from which it was derived). Six types are used: Identity (demographics, profession, education, relationships); Preference (likes and habits); Goal (plans and upcoming events); Interest (hobbies and topics); Activity (recurring or past actions); and Event (situations or milestones that affect decisions). Extraction and each write are performed by an LLM; memory is updated once every few sessions to limit cost.

\vspace{-0.2cm}
\paragraph{Memory Upsert} For the first conversation, all extracted facts are treated as new and receive new identifiers. For subsequent conversations, extracted facts are upserted \ie only new facts are inserted while pre-existing ones are updated if necessary. 

\vspace{-0.2cm}
\paragraph{Memory Consolidation.} When the new fact count reaches a threshold (50 new facts since the last merge), facts are merged based on semantic similarity and temporal proximity. Fact texts are embedded and clustered by cosine similarity; within each cluster, only facts whose most recent date falls within a fixed time window (7 days) are merged. An LLM consolidates each such group into a single fact; provenance is aggregated and the original identifiers are retained.

\vspace{-0.2cm}
\paragraph{Retrieval.} At query time, retrieval is performed over the memory facts by embedding-based similarity (like a RAG); PGR applies its prospective loop on top of this interface. 

\noindent
All the prompts used for extraction, update, and merge are provided in Appendix~\ref{sec:appendix}.

\subsection{Prospection Guided Retrieval}
\label{sec:pgr-retrieval}

The core of PGR is an iterative \emph{simulate--retrieve--refine} process that mirrors human prospection. Rather than treating a user query as a static search term, PGR treats it as a starting point for mental simulation. The underlying intuition is that to find a memory that is logically relevant but semantically distant, the system must first ``imagine'' the future scenarios where that memory would be required. 

This mechanism acts as a dynamic dialogue between a generative planner (the PFC analog) and the memory substrate (the hippocampal analog). The process begins with a query augmentation step to ground the user intent, followed by two phases of prospection that iteratively retrieve and refine relevant memories:

\squishlist
    \item \textbf{Initial Prospection (Phase 1):} The system generates a broad ``horizon'' of potential future actions. This simulation can be operationalized as either a linear \textit{Chain-of-Thought (CoT)} or a branching \textit{Tree-of-Thought (ToT)} structure. Each hypothesized step serves as an independent \textbf{sub-query} to the memory store, potentially surfacing latent facts that the original query alone would fail to trigger. 
    \item \textbf{Personalized Prospection (Phase 2):} The system refines the simulation by conditioning it on the facts retrieved in Phase 1. This personalization allows the simulation to reflect user-specific constraints, proactively retrieving memories that match the refined context.
\squishend

This iterative loop ensures that each retrieved memory informs a more accurate simulation of the future, which in turn serves as a more precise probe for deeper, latent memories. 

\paragraph{Formal Setup.}
Let $\mathcal{M}$ denote the memory store. 
The memory substrate is defined via an abstract retrieval function $\phi(x; \mathcal{M}, \Theta)$, returning a set of facts relevant to a sub-query $x$. In our reference implementation, $\Theta = \{K, \tau\}$, representing the top-$K$ facts above semantic similarity threshold $\tau$. For different memory architectures, $\Theta$ may instead encode traversal depth, temporal windows, or community hierarchy. We denote $R$ as the set of accumulated facts, initially $R = \emptyset$.

\paragraph{Query Augmentation.} 
PGR first generates an augmented query $q_A$ to resolve underspecified user queries by retrieving tightly relevant context:
\vspace{-.8em}
\begin{equation}
q_A = \mathrm{LLM}(P_A(q,\, \phi(q;\mathcal{M}, \Theta_o)))
\end{equation}

where $P_A$ instructs the model to disambiguate using context $\Theta_o$ (e.g., $K_o, \tau_o$). This ensures the simulation begins with a high-fidelity representation of user intent.

\paragraph{Phase 1: Initial Prospection.} 
The LLM processes $q_A$ to produce a prospection sequence $S^0 = \{s^0_1, \ldots, s^0_L\}$ of hypothesized subgoals or future contexts:
\vspace{-.8em}
\begin{equation}
S^0 = \mathrm{LLM}(P_{\mathrm{sim}}(q_A))
\end{equation}

Each $s^0_l \in S^0$ is used as a sub-query to the memory store, and retrieved facts are merged with the initial query results:
\vspace{-.8em}
\begin{equation}
R \leftarrow \phi(q;\mathcal{M}, \Theta) \cup \bigcup_{l=1}^{L} \phi(s^0_l;\mathcal{M}, \Theta)
\end{equation}

\paragraph{Phase 2: Personalized Prospection.} 
Using the accumulated knowledge $R$, the LLM iteratively refines the prospection sequence, making it personalized to the user:
\begin{equation}
S^i = \mathrm{LLM}(P_{\mathrm{ref}}(q_A, R, S^{i-1}))
\end{equation}

For each refined sub-query $s^i_l \in S^i$, similar facts are retrieved. $\Delta^i$ denotes the set of unique, previously unseen facts discovered during iteration $i$:
\vspace{-.8em}
\begin{equation}
\Delta^i = \bigcup_l \phi(s^i_l;\mathcal{M}, \Theta) \setminus R
\end{equation}

The memory set is updated $R \leftarrow R \cup \Delta^i$, and the loop terminates when $|\Delta^i|$ falls below a convergence threshold $\delta$ or a maximum iteration $I_{\max}$ is reached. The final set $R^*$ is returned alongside a prospection reasoning summary detailing how the simulation guided memory recall.

\begin{algorithm}[t]
\caption{Prospection-Guided Retrieval (PGR)}
\label{alg:pgr}
\begin{algorithmic}[1]
\Require Query $q$, memory store $\mathcal{M}$, abstract retrieval function $\phi$, augmentation parameters $\Theta_o$, retrieval parameters $\Theta$, convergence threshold $\delta$, max iterations $I_{\max}$
\Ensure Final retrieved fact set $R^*$

\State $R \leftarrow \emptyset$
\State \Comment{\textit{Query Augmentation}}
\State $q_A \leftarrow \mathrm{LLM}(P_A(q,\; \phi(q;\mathcal{M}, \Theta_o)))$

\State \Comment{\textit{Phase 1: Initial Prospection}}
\State $S^0 \leftarrow \mathrm{LLM}(P_{\mathrm{sim}}(q_A))$
\State $R \leftarrow \phi(q;\mathcal{M}, \Theta) \;\cup \displaystyle\bigcup_{l=1}^{|S^0|} \phi(s^0_l;\mathcal{M}, \Theta)$

\State \Comment{\textit{Phase 2: Personalized Prospection}}
\For{$i = 1, 2, \ldots, I_{\max}$}
    \State $S^i \leftarrow \mathrm{LLM}(P_{\mathrm{ref}}(q_A,\, R,\, S^{i-1}))$
    \State $\Delta^i \leftarrow \displaystyle\bigcup_{l=1}^{|S^i|} \phi(s^i_l;\mathcal{M}, \Theta) \;\setminus\; R$
    \State $R \leftarrow R \cup \Delta^i$
    \If{$|\Delta^i| < \delta$} \textbf{break} \EndIf
\EndFor
\State \Return $R^* \leftarrow R$
\end{algorithmic}
\end{algorithm}

%% file: benchmark.tex
\section{Benchmark Dataset: MemoryQuest}
\label{sec:dataset}

The need for \textbf{MemoryQuest} stems from the complexity of real-world assistant usage, where interactions span months and disparate tasks like travel booking or event planning. In these settings, a single query often depends on a ``thread'' of chronologically consistent facts that evade simple semantic search. Most existing benchmarks are limited to single sessions or lack logical evolution over time. Furthermore, retrieval tasks are often ``semantically easy,'' allowing standard RAG to succeed via keyword overlap. \textbf{MemoryQuest} provides a necessary stress-test for long-term memory, requiring the synthesis of multiple, semantically distant memories buried across a multi-domain history.

\subsection{Gap Analysis and Comparison}
Prior datasets target complementary aspects of long-term memory but lack the combination of temporal grounding and semantic difficulty required for proactive retrieval:

\squishlist
    \item \textbf{Single Session or Temporal Inconsistency:} \emph{LongMemEval}~\cite{wu2024longmemevalbenchmarkingchatassistants} and \emph{MSC}~\cite{xu-etal-2022-beyond} focus on evidence within continuous logs. Unlike \emph{LoCoMo}~\cite{maharana2024evaluating}, which uses event graphs, or \emph{MSC}, which lacks explicit dates, \emph{MemoryQuest} provide a dated, multi-session history requiring retrieval of $3$--$5$ distinct memories spread across disparate task domains.
    \item \textbf{Semantic Retrieval Bias:} In \emph{PersonaMem}~\cite{jiang2025personamem} and \emph{ImplexConv}~\cite{li2025implexconv}, evidence is often semantically similar to the query. \emph{MemoryQuest} enforces a \textbf{similarity bottleneck}, pairing queries with required facts filtered for low semantic overlap.
    \item \textbf{Lack of Fine-Grained Evidence:} \emph{MemoryQuest} provides dated, atomic facts. This enables precise, independent evaluation of both the \textbf{retrieval policy} and the \textbf{generation model}.
\squishend

\subsection{Dataset Construction Pipeline}
MemoryQuest is built on \emph{PersonaLens} profiles~\cite{zhao-etal-2025-personalens}, providing demographics across 20 domains. We generate \emph{in-situ} queries involving planning and follow-through rather than standalone factoid prompts via a three-step pipeline:

\paragraph{Step 1: Query Generation.} 
An LLM generates candidate queries, each including a query date, reasoning, and 3--5 \textbf{required references}. We retain candidates with average query--reference cosine similarity $\gamma \le 0.3$ and keep at most $n_q \le 15$ queries per user, prioritizing those with the lowest similarity scores.

\paragraph{Step 2: Timeline Synthesis.} 
For each query, an LLM builds a chronologically ascending timeline of 5--10 events, starting shortly before the earliest required-reference date. Each required reference is embedded in exactly one timeline event, while remaining events are persona-grounded \textbf{filler} chats for a realistic multi-domain conversation history.

\paragraph{Step 3: Dialogue Expansion.} 
Each timeline event is expanded into a multi-turn conversation (5--30 turns) via an LLM, conditioned on user demographics and domain summaries for logical continuity. A separate memory-extraction step (Section~\ref{sec:pgr-memory}) converts these logs into the structured fact store used at test time.

\paragraph{Scope and Generalizability.}
MemoryQuest is intentionally designed as a challenging stress-test for semantically indirect long-term personalization. The $\gamma \le 0.3$ filter selects queries whose required references have low embedding similarity to the query, creating a setting where direct single-query retrieval is expected to be challenging. We therefore do not present MemoryQuest as representative of all retrieval workloads; rather, it targets an important subset of assistant interactions: under-specified requests that require recovering temporally distributed, personalized evidence from long histories. To assess whether PGR's benefits extend beyond this constructed setting, we also evaluate on ImplexConv \cite{li2025implexconv} and PersonaMem \cite{jiang2025personamem}, two independently published benchmarks with different construction procedures and task objectives, where PGR improves retrieval recall over all baselines and achieves strong response-level win rates.

%% file: evaluation.tex
\section{Experiments}
\label{sec:evaluation}

We evaluate PGR against multiple retrieval and reasoning baselines on long-term, multi-session conversational tasks to test whether prospective, simulation-driven retrieval provides more anticipatory and contextually grounded personalization than conventional retrospective retrieval. We use three benchmarks. The \textbf{MemoryQuest} benchmark (Section~\ref{sec:dataset}) includes 50 users and \textbf{535 queries} across 20 task domains; each query requires \textbf{3--5} references, which serve as retrieval and answer ground truth. Users have an average of \textbf{77.6 sessions} with \textbf{6.2 exchanges per session}. {ImplexConv}~\cite{li2025implexconv} provides long multi-session dialogues with opposed and supportive implicit-reasoning scenarios; we use only the \textit{opposed} setting, where relevant context is semantically distant and contradicts the surface persona. Each query has a single required reference; we evaluate \textbf{196 queries} across 85 users (sampled from 1,550 users in the full dataset), with an average of \textbf{110 sessions} and \textbf{5.8 exchanges per session}. {PersonaMem}~\cite{jiang2025personamem} provides persona profiles and long conversation histories; we use the 32k chat-history variant and sample \textbf{50 random profiles}, excluding \texttt{ask\_to\_forget} and \texttt{sensitive\_info} queries to focus on preference-grounded personalization, yielding \textbf{894 queries} (sampled from 200 users and 3,441 queries in the full dataset); each user has a single long chat history averaging \textbf{116 exchanges} ($\sim$32k tokens). Overall, we evaluate 1,625 queries across 185 user profiles. We release the MemoryQuest generation framework to support reproducibility and scaling to 1,500+ user profiles.

\subsection{Baselines}
We compare PGR against three other state of the art memory retrieval baselines. Experiments are run using Azure OpenAI GPT-4o for simulation and answer generation, and \texttt{text-embedding-3-small} for embedding and retrieval \cite{openai2024embeddings}. Additional experiments using PGR-TOT with DeepSeek-V3.2 as the generation model are reported in Appendix~\ref{sec:deepseek_results}.



\textbf{Mem0$^\texttt{g}$.} \cite{chhikara2025mem0} utilizes a Neo4j-backed entity-relation graph to store memories. Retrieval combines vector search with graph relation extraction to test if explicit relational structures can compensate for the absence of prospection. We evaluate at $K=20$.

\textbf{GraphRAG.} \cite{edge2024graphrag} indexes dialogue sessions into an entity graph and retrieves the top-$K=5$ community reports via embedding similarity to evaluate structured graph summarization as a non-prospective alternative.

\textbf{TaciTree.} \cite{li2025implexconv} utilizes a hierarchical tree for implicit reasoning retrieval. Facts are clustered via UMAP and Gaussian Mixture Models (max size $k=6$), then summarized by an LLM to form leaf nodes. The tree is built iteratively until the root contains $L=15$ clusters. At inference, an LLM performs top-down pruning to select the top-$K=5$ most relevant leaf summaries, avoiding exhaustive search.

\paragraph{Hyperparameters.} For all PGR variants, we set the convergence threshold to $\delta = 5$ newly retrieved facts and the maximum number of refinement iterations to $I_{\max} = 10$. Query-only baselines retrieve the top $K = 20$ facts; each PGR sub-query retrieves the top $K = 5$ facts with embedding similarity threshold $\tau = 0.3$. For query augmentation in Phase 2, we apply a stricter threshold $\tau_o = 0.6$ to ensure only high-confidence context informs refinement. These settings are fixed across all datasets.

\subsection{Evaluation Metrics}

We use Azure OpenAI GPT 5.2 for evaluating the responses. Prompts are provided in the Appendix. 

\begin{wrapfigure}{r}{0.44\textwidth} 
    \centering
    \vspace{-10pt} 
    \includegraphics[width=0.42\textwidth]{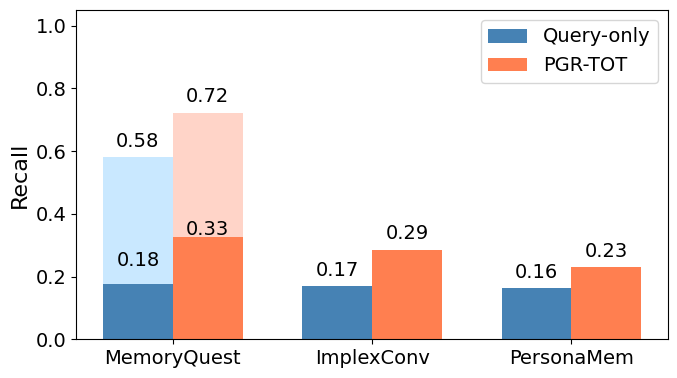}
    \caption{\small{Results comparing Retrieval Recall using Query-only and PGR-TOT. For MemoryQuest, lighter bars: Recall, and darker bars: Recall$_{Exact}$.}}
    \label{fig:rag-vs-pgr}
    \vspace{-12pt} 
\end{wrapfigure}

\paragraph{Retrieval-level:}
For each method, we obtain the \emph{retrieved context} (e.g., the set of facts or chat excerpts retrieved to generate the answer). The LLM is given the query, the list of required references, and the retrieved context, with a simple instruction to independently identify (binary, Yes/No) whether each reference is found explicitly or implicitly. We report \textbf{Recall} as the fraction of required references judged present, averaged over queries. For ImplexConv and PersonaMem, the judgment is binary. Since MemoryQuest has multiple references, we report the fraction as well as \textbf{Recall$_{Exact}$}, which is 1 if all references are identified, else 0.

\paragraph{Response-level:}
We also compare the \emph{final answers} generated using PGR and each baseline by constructing an LLM-based user simulator that has knowledge of all facts about the user context. For each query, the two responses are presented to the simulated user, along with the list of required references. The output indicates which response is more useful and proactive (acknowledging past context, anticipating needs, and giving actionable guidance) or marks a tie. We repeat this experiment twice for each query by reversing the order of responses to avoid position bias and take the average of both. Prompts are provided in the Appendix. We report \textbf{Win Rate} (\%) of PGR versus each baseline as the fraction of queries where the simulated user prefers the PGR response. We also conducted a human study to evaluate the efficacy of LLM-as-a-Judge on a smaller subset of queries, comprising approximately 4\% of MemoryQuest, 5\% of ImplexConv, and 2\% of PersonaMem queries. Human evaluations were conducted on each PGR-baseline pair, and results are reported in Table \ref{tab:main_results}. Experiments were conducted on Amazon MTurk. IRB exemption was received for this study.

\section{Results and Discussion}
\label{sec:results}

\begin{table}[t]
\centering
\setlength{\tabcolsep}{4pt}
\caption{Average Recall and Win Rate(\%) of PGR-TOT over baseline methods evaluated on MemoryQuest, ImplexConv, and PersonaMem datasets. Human* evaluations were done on a subset of queries. Higher is better.}
\vspace{1pt}
\begin{tabular}{l|ccccc|ccc|ccc}
\toprule
\textbf{Dataset $\rightarrow$} & \multicolumn{5}{c|}{\textbf{MemoryQuest}} 
& \multicolumn{3}{c|}{\textbf{ImplexConv}} 
& \multicolumn{3}{c}{\textbf{PersonaMem}} \\
\cmidrule(lr){2-6} \cmidrule(lr){7-9} \cmidrule(lr){10-12}
\textbf{Method $\downarrow$} & Recall & Recall$_{Exact}$ & \multicolumn{2}{c}{Win Rate$_{PGR}$} & & Recall & \multicolumn{2}{c}{Win Rate$_{PGR}$} & Recall & \multicolumn{2}{c}{Win Rate$_{PGR}$} \\
\cmidrule(lr){4-5} \cmidrule(lr){8-9} \cmidrule(lr){11-12}
& & & \small LLM & \small Humans$^*$ & & & \small LLM & \small Humans$^*$ & & \small LLM & \small Humans$^*$ \\
\midrule
GraphRAG & \med{0.160} & \low{0.003} & \vhigh{96.6} & \high{90.0} & & \low{0.072} & \vhigh{96.2} & \high{70.0} & \low{0.011} & \vhigh{98.4} & \high{80.0} \\
TaciTree & \med{0.235} & \low{0.024} & \vhigh{94.3} & \high{75.0} & & \low{0.122} & \high{89.9} & \med{60.0} & \low{0.073} & \high{88.7} & \med{65.0} \\
Mem0$^{\texttt{g}}$ & \med{0.256} & \low{0.015} & \vhigh{94.7} & \vhigh{95.0} & & \low{0.050} & \vhigh{93.5} & \med{70.0} & \low{0.121} & \vhigh{92.9} & \med{65.0} \\
\textbf{PGR-TOT} & \vhigh{\textbf{0.723}} & \high{\textbf{0.326}} & - & - & & \med{\textbf{0.286}} & - & - & \med{\textbf{0.229}} & - & - \\
\bottomrule
\end{tabular}
\label{tab:main_results}
\end{table}

Results in Table~\ref{tab:main_results}, demonstrate that Prospection-Guided Retrieval (PGR) outperforms traditional retrieval baselines across all evaluated datasets and metrics.

\paragraph{Higher Retrieval Recall.} 
On the MemoryQuest benchmark, PGR-TOT achieves a recall of \textbf{0.723}, nearly tripling the performance of the strongest baseline, Mem0 (0.256), and vastly exceeding GraphRAG (0.160). This trend persists across ImplexConv and PersonaMem, where PGR consistently surfaces relevant context that passive semantic search fails to identify. These gains are particularly pronounced in \textbf{Recall$_{Exact}$} metrics (\textbf{0.326} vs.\ 0.015 for Mem0), indicating that PGR is uniquely capable of retrieving the complete "thread" of facts required for complex reasoning.

Figure \ref{fig:rag-vs-pgr} compares retrieval performance between a standard \textit{Query-only} baseline and PGR-TOT. We decompose these results into Recall$_{Exact}$ (dark regions)—representing queries where the entire ground-truth set is recovered—and fractional recall (light regions).

The visualization underscores the "prospection gap": while the Query-only baseline struggles with a total recall of only \textbf{0.58} on MemoryQuest, PGR-TOT achieves \textbf{0.723}.  This demonstrates that by simulating future trajectories, PGR effectively bypasses the semantic similarity bottleneck, successfully reconstructing the complete "thread" of chronologically distant facts required for complex personalization.

\begin{wraptable}{r}{0.45\textwidth}
\centering
\renewcommand{\arraystretch}{1.2}
\setlength{\tabcolsep}{4pt} 
\vspace{-12pt}
\caption{Aggregated comparison matrix between judgments of humans (rows) and LLM (columns).}
\small
\begin{tabular}{l|>{\centering\arraybackslash}p{1.1cm}>{\centering\arraybackslash}p{1.1cm}>{\centering\arraybackslash}p{1.1cm}}
\toprule
\textbf{Human $\backslash$ LLM} & \textbf{PGR} & \textbf{Tie} & \textbf{Baseline} \\
\midrule
\textbf{PGR} & \cellcolor{pink!70} 70.0 & \cellcolor{pink!20} 4.0 & \cellcolor{pink!0} 2.0 \\
\textbf{Tie} & \cellcolor{pink!20} 8.0 & \cellcolor{pink!0} 0.0 & \cellcolor{pink!10} 0.7 \\
\textbf{Baseline} & \cellcolor{pink!40} 12.0 & \cellcolor{pink!20} 2.0 & \cellcolor{pink!20} 1.3 \\
\bottomrule
\end{tabular}
\label{tab:human-llm-matrix}
\vspace{-0pt}
\end{wraptable}

\paragraph{High-Quality Response Generation.} 
The improvement in retrieval translates directly to superior response quality. As shown in Table~\ref{tab:main_results}, PGR achieves dominant win rates in LLM-as-a-judge evaluations, ranging from \textbf{88.7\% to 98.4\%}. Crucially, human evaluations confirm this preference, with annotators favoring PGR responses in \textbf{60--95\%} of cases. Table~\ref{tab:human-llm-matrix} reports the aggregated human-vs-LLM agreement matrix across all baseline pairs and datasets; humans and the LLM judge agree on preferring PGR in \textbf{70\%} of all evaluated cases. Human annotators noted that PGR-based responses were significantly more proactive, correctly utilizing long-term personal history to provide actionable guidance.

The empirical success of PGR highlights a structural limitation in contemporary ``lookup-based'' memory architectures. By decoupling memory retrieval from underlying storage and implementing an iterative \textit{simulate $\rightarrow$ retrieve $\rightarrow$ refine} cycle, PGR enables the system to proactively anticipate informational needs. 
\begin{wraptable}{r}{0.45\textwidth}
    \centering
    \setlength{\tabcolsep}{4pt}
    \vspace{-.5em} 
    \caption{Effect of prospection strategies.}
    \small
    \begin{tabular}{l|cc}
        \toprule
        \textbf{PGR Method} & \textbf{Recall} & \textbf{Win Rate (\%)} \\
        \midrule
        TOT \textit{vs} COT  & \high{0.723 \textit{\textbf{>}} 0.718} & \low{55.67} \\
        \bottomrule
    \end{tabular}
    \label{tab:ablation-tot-cot}
    \vspace{-1em}
\end{wraptable}
This mechanism facilitates the identification of ``latent'' memories: contextual facts that, while lacking immediate semantic proximity to the current dialogue turn, are crucial for achieving the user's long-term objectives. Our findings suggest that \textit{prospection} is a key paradigm for developing highly personalized, context-aware conversational agents.

\begin{table}[h]
    \centering
    \setlength{\tabcolsep}{3pt} 
    
    \begin{minipage}{0.49\textwidth}
        \centering
        \caption{Effect of iterative prospection.}
        \begin{tabular}{l|cc|c}
            \toprule
            \textbf{Method} & \multicolumn{2}{c}{\textbf{Recall}} & \textbf{Win Rate} \\
             & \textbf{Base} & \textbf{Iterative} & \textbf{Iterative} \\
            \midrule
            PGR-TOT & \med{0.677} & \high{0.723} & \high{70.13} \\
            PGR-COT & \med{0.645} & \high{0.718} & \high{75.41} \\
            \bottomrule
        \end{tabular}
        \label{tab:ablation-iterative}
    \end{minipage}
    \begin{minipage}{0.49\textwidth}
        \centering
        \caption{Effect of prospection summary.}
        \begin{tabular}{l|cc|c}
            \toprule
            \textbf{Method} & \multicolumn{2}{c}{\textbf{Recall}} & \textbf{Win Rate} \\
             & \textbf{$\sim$PAG} & \textbf{PAG} & \textbf{PAG} \\
            \midrule
            PGR-TOT & \med{0.703} & \high{0.722} &  \high{75.73} \\
            PGR-COT & \high{0.722} & \high{0.735} &  \high{74.37} \\
            \bottomrule
        \end{tabular}
        \label{tab:ablation-prospection}
    \end{minipage}
\end{table}

We conduct ablations on MemoryQuest to isolate the contributions of iterative retrieval, the use of the prospection summary in answer generation, and the type of prospection strategy within PGR.

\textbf{Iterative Prospection} consistently improves memory recall and answer quality. As shown in Table~\ref{tab:ablation-iterative}, both PGR-COT and PGR-TOT achieve higher recall and win rates when sub-queries are retrieved and refined iteratively.

\textbf{Prospection-Augmented Generation (PAG).} Conditioning answer generation on the prospection summary further boosts performance. Table~\ref{tab:ablation-prospection} shows that including PAG improves recall and win rate over using retrieved facts alone for both COT and TOT variants.

\textbf{Prospection Strategy.} Comparing PGR-TOT and PGR-COT (Table~\ref{tab:ablation-tot-cot}) reveals that branching tree-of-thought exploration slightly outperforms linear chain-of-thought in Recall, though win rate differences are modest. 

Overall, the results indicate that iterative retrieval, prospection-guided generation, and the choice of prospection strategy each contribute to more effective memory retrieval and response generation in PGR.

\textbf{Limitations.}
PGR’s iterative process improves accuracy by carefully planning and exploring potential future trajectories, but this comes at the cost of additional LLM calls compared to single-pass retrieval. While PGR remains cheaper than TaciTree per query (Appendix Table~\ref{tab:llm-budget}), the multi-phase prospection can be resource-intensive for real-time applications. Future work could reduce this overhead by training specialized lightweight retrievers on the generated prospection traces, enabling faster memory selection while retaining much of the anticipatory reasoning benefit. 

%% file: conclusion.tex
\section{Conclusion}
\label{sec:conclusion}

We introduced Prospection-Guided Retrieval (PGR), a retrieval policy that is independent of the underlying memory substrate. Instead of matching only on retrospective similarity to the current turn, PGR runs an iterative \emph{simulate $\rightarrow$ retrieve $\rightarrow$ refine} loop so that simulated futures surface latent and counterfactual facts that embedding or graph search often misses. This addresses a recurring limitation in standard RAG and GraphRAG pipelines, where retrieval is largely fixed by the index structure and behaves like a narrow lookup even when users need long-horizon, goal-directed assistance. We also introduced MemoryQuest, a multi-session benchmark that requires coordinating several dated references under low query--reference embedding similarity, and we evaluated PGR on MemoryQuest, ImplexConv, and PersonaMem. Across these datasets, PGR improves recall (including full multi-reference recovery) and produces answers preferred by both LLM judges and human annotators. Overall, these results support prospection as a complementary retrieval strategy when an assistant must assemble scattered personal context across sessions rather than return the closest passage to a single query.

%% file: appendix.tex
\subsection{PGR with Alternative Generation Models}
\label{sec:deepseek_results}

To assess whether PGR's retrieval gains are robust to the choice of generation model, we run PGR-TOT with DeepSeek-V3.2 replacing GPT-4o for prospection and answer generation, while keeping the same retrieval pipeline, embeddings, and hyperparameters. Table~\ref{tab:deepseek_results} reports retrieval Recall and Recall$_{Exact}$ (MemoryQuest only) for the base single-phase and iterative two-phase variants across all three datasets. Results are consistent with the main GPT-4o findings, confirming that PGR's gains are not model-specific.

\begin{table}[h]
\centering
\setlength{\tabcolsep}{6pt}
\caption{Recall of PGR-TOT (DeepSeek-V3.2) across datasets. Base = single-phase prospection; Iterative = two-phase. Recall$_{Exact}$ reported for MemoryQuest only.}
\vspace{1pt}
\begin{tabular}{l|cc|c|c}
\toprule
\textbf{Dataset} & \textbf{Base Recall} & \textbf{Iterative Recall} & \textbf{Base Recall$_{Exact}$} & \textbf{Iterative Recall$_{Exact}$} \\
\midrule
MemoryQuest & \med{0.671} & \vhigh{\textbf{0.748}} & \med{0.245} & \high{\textbf{0.348}} \\
ImplexConv  & \med{0.308} & \high{\textbf{0.374}}  & --          & --                   \\
PersonaMem  & \med{0.210} & \med{\textbf{0.229}}   & --          & --                   \\
\bottomrule
\end{tabular}
\label{tab:deepseek_results}
\end{table}

Iterative prospection consistently improves recall over the single-phase base across all three datasets, mirroring the pattern observed with GPT-4o (Table~\ref{tab:main_results}). The gains on MemoryQuest (\textbf{0.748} Recall, \textbf{0.348} Recall$_{Exact}$) exceed those of the GPT-4o variant (0.723 / 0.326), while ImplexConv and PersonaMem show the same directional improvement. This confirms that PGR's prospection-guided retrieval framework generalizes across generation models.

\vspace{1em}

\subsection{Computational Cost.}
Table~\ref{tab:llm-budget} reports the average number of LLM calls and unique facts retrieved per query for PGR-TOT and each baseline. PGR-TOT requires approximately 4--5 LLM calls per query (one Phase-1 prospection + avg.\ Phase-2 refinements + one prospection summary + one answer generation), compared to 1 for GraphRAG and $\sim$2 for Mem0$^\texttt{g}$. TaciTree’s higher count reflects its level-by-level LLM pruning at inference (offline tree-construction calls excluded). The embedding-based retrieval sub-steps are parallelizable and add negligible latency relative to the LLM calls.

\begin{table}[h]
\centering
\setlength{\tabcolsep}{5pt}
\renewcommand{\arraystretch}{1.1}
\caption{Average query-time LLM calls and unique facts retrieved per query for PGR-TOT.}
\vspace{2pt}
\small
\begin{tabular}{l|cccc|cc}
\toprule
\textbf{Dataset} & \textbf{GraphRAG} & \textbf{Mem0$^\texttt{g}$} & \textbf{TaciTree} & \textbf{PGR-TOT} & \textbf{PGR Ph-1 Facts} & \textbf{PGR Ph-2 New Facts} \\
\midrule
MemoryQuest & 1.0 & 2.0 & $\sim$13.2 & \textbf{4.79} & 26.78 & 8.41 \\
ImplexConv  & 1.0 & 2.0 & $\sim$15.9 & \textbf{4.32} & 22.41 & 3.85 \\
PersonaMem  & 1.0 & 2.0 & $\sim$9.7  & \textbf{4.41} & 13.04 & 4.83 \\
\bottomrule
\end{tabular}
\label{tab:llm-budget}
\end{table}


\subsection{End-to-End PGR Trace: MemoryQuest Example}
\label{sec:e2e-example}

The following is a complete PGR-TOT trace on a MemoryQuest query where all four required references were successfully retrieved. The system runs Phase~1 (initial TOT, 15 facts retrieved), then two iterative refinement phases that progressively personalize the tree and surface new facts. Notice how the node constraints evolve across iterations as retrieved facts ground the simulation in the user's specific context.

\vspace{0.5em}
\noindent\textbf{Query, Date, and Required References}
\begin{lstlisting}[style=appendixbox]
Date:  2026-03-05
Query: "I want to pre-order a new game that just got announced.
        Check if it makes sense for me to do it now."

Required references (all 4 retrieved by PGR-TOT):
  [1] GPU overheating issue -- user plans to reduce overheating during gaming (2026-02-20).
  [2] March trip funds locked -- user frames trip money as 'already spent' (2026-02-23).
  [3] Upcoming online course expenses constraining GPU/hardware budget (2026-03-03).
  [4] Backlog of unfinished games -- user committed to completing RDR2 before any new
      title (2026-03-01).
\end{lstlisting}

\vspace{0.5em}
\noindent\textbf{Phase 1 --- Initial TOT (initial, pre-retrieval)}
\begin{lstlisting}[style=appendixbox]
A1  Check the game's release date and pre-order availability
    constraints: official announcement, platform compatibility, region restrictions
    +-- A2  Look up gameplay trailers or developer previews
    |       constraints: video quality, gameplay mechanics, genre preference
    |       +-- A4  Read reviews from early access / beta testers
    |       |       constraints: reviewer credibility, performance feedback
    |       |       +-- A7  Decide if game aligns with personal gaming preferences
    |       |               constraints: time available, interest in story/mechanics
    |       |               +-- A9  [leaf] Pre-order or wait for post-release reviews
    |       |                       constraints: confidence in purchase, willingness to wait, FOMO
    |       +-- A5  [leaf] Compare with similar games in the genre
    |               constraints: gameplay depth, replay value, prior genre experience
    +-- A3  Check pre-order bonuses or exclusive content
            constraints: bonus relevance, edition types, availability
            +-- A6  Check the pre-order price and payment options
                    constraints: budget, refund policy, currency exchange rates
                    +-- A8  [leaf] Evaluate risk of bugs / incomplete features at launch
                            constraints: developer track record, past launch issues, patch history
\end{lstlisting}

\vspace{0.5em}
\noindent\textbf{Phase 1 --- Retrieved Facts} {\small(all 15 facts passed to Iteration~1)}
\begin{lstlisting}[style=appendixbox]
[2026-03-01] [Preference] User prefers finishing one main game at a time; avoids starting
                          new games until current one is completed or consciously dropped.
[2026-02-23] [Preference] 2-rule purchase system: under EUR 10 requires 24-hr wait;
                          over EUR 10 must fit fun money AND be played immediately.
                          User now muting sale notifications until after March trip.
[2026-02-23] [Preference] Frames trip money as 'already spent' to mentally lock it.
                          Rule: "Trip money is locked. Games from fun money only.
                          If fun money = 0 EUR, wishlist and wait."
[2026-02-23] [Goal]       Fun money for March: EUR 10 (tight) to EUR 15-20 (normal);
                          trip savings locked in separate pocket labelled "DO NOT TOUCH".
[2026-03-01] [Goal]       Prioritising RDR2 main story: 90-min capped sessions x2/week
                          (Wed afternoons). Witcher 3 parked guilt-free.
[2026-03-01] [Preference] Gaming sessions capped at 90 min to prevent late nights and
                          maintain focus for job applications and 9:00 AM alarm.
[2025-12-29] [Preference] Keyboard and mouse for gaming; mostly RPGs; prefers long sessions.
[2026-03-03] [Preference] Prefers maintenance over upgrades for GTX 1660 Super GPU due to
                          tight budget, locked travel savings, upcoming online course expenses.
[2026-02-21] [Goal]       Tracks fun money (EUR 15-20) weekly; Sunday 18:00 reminder
                          "Quick check: fun money (5 min)".
[2025-12-29] [Preference] Enjoys gaming evenings/weekends to save money but limits
                          bedtime shifts to avoid disrupting weekday routines.
[2026-01-29] [Goal]       EUR 50 bill as visual safety net for next 3 days -- keep untouched
                          as a measure of financial control.
[2026-02-14] [Preference] Prefers fun sci-fi or action movies for casual friend hangouts.
[2025-12-27] [Preference] Budget-friendly charger replacements; uses Amazon.es, PcComponentes.
[2026-01-16] [Goal]       Calendar note for Friday interviews: "Possible transport disruption
                          -- message recruiter by 7:30 if issues".
[2026-02-03] [Goal]       Recurring 'Available for interviews' note protecting early afternoons.
\end{lstlisting}

\vspace{0.5em}
\noindent\textbf{Iteration 1 --- Personalized TOT} {\small(9 nodes; constraints updated to reflect retrieved facts)}
\begin{lstlisting}[style=appendixbox]
A1  Check the game's release date and pre-order availability
    constraints: official announcement, platform compatibility, region restrictions
    +-- A2  Look up gameplay trailers or developer previews
    |       constraints: genre preference, [+relevance to current gaming interests, RPGs]
    |       +-- A4  Read reviews from early access / beta testers
    |       |       constraints: reviewer credibility, [+alignment with gaming preferences]
    |       |       +-- A7  Decide if game aligns with personal gaming preferences
    |       |               constraints: time available, [+current RDR2 commitment --
    |       |                            must finish before starting new games]
    |       |               +-- A9  [leaf] Pre-order or wait
    |       |                       constraints: FOMO, [+fun money only; current RDR2 focus]
    |       +-- A5  [leaf] Compare with similar games
    |               constraints: gameplay depth, [+preference for RPGs and long sessions]
    +-- A3  Check pre-order bonuses
            +-- A6  Check pre-order price and payment options
                    constraints: budget, [+fun money only; >EUR 10 requires immediate
                                  playability; refund policy]
                    +-- A8  [leaf] Evaluate risk of bugs at launch
                            constraints: track record, [+preference for maintenance over
                                         upgrades; tight budget and online course expenses]

New facts retrieved in Iteration 1 (5 new facts):
  [2026-02-21] [Goal]  Track fun money (EUR 15-20) weekly; Sunday 18:00 reminder.
  [2026-01-29] [Goal]  EUR 50 bill as visual safety net -- keep untouched.
  [2026-02-20] [Goal]  Plans to open PC case side panel during gaming sessions to
                       test airflow and reduce GPU overheating.
  [2025-12-29] [Event] Eye strain during late-night gaming; uses blue light filters,
                       reduced brightness, warm lighting to mitigate fatigue.
  [2026-02-23] [Goal]  Weekly Sunday reminder "March trip fund still locked" to
                       reinforce saving discipline.
\end{lstlisting}

\vspace{0.5em}
\noindent\textbf{Iteration 2 --- Further Refined TOT} {\small(9 nodes; constraints tightened with Iteration~1 facts)}
\begin{lstlisting}[style=appendixbox]
A1  Check the game's release date and pre-order availability
    constraints: [+release date relevance to March trip AND RDR2 completion goal]
    +-- A2  Look up gameplay trailers or developer previews
    |       constraints: genre preference, RPGs, [+alignment with 90-min session cap]
    |       +-- A4  Read reviews from early access / beta testers
    |       |       constraints: reviewer credibility, [+focus on avoiding games with
    |       |                    launch bugs or incomplete features]
    |       |       +-- A7  Decide if game aligns with personal gaming preferences
    |       |               constraints: [+RDR2 in progress; 90-min sessions x2/week;
    |       |                            starting new game violates current commitment]
    |       |               +-- A9  [leaf] Pre-order or wait
    |       |                       constraints: [+fun money EUR 15-20 for March;
    |       |                                    trip money non-negotiable; FOMO low]
    |       +-- A5  [leaf] Compare with similar games
    |               constraints: [+RDR2/Witcher 3 backlog already present]
    +-- A3  Check pre-order bonuses
            +-- A6  Check pre-order price and payment options
                    constraints: [+fun money capped EUR 15-20; >EUR 10 requires
                                  immediate playability -- not possible while on RDR2]
                    +-- A8  [leaf] Evaluate risk of bugs at launch
                            constraints: [+GPU overheating issue; prefers maintenance;
                                         online course costs in March]

New facts retrieved in Iteration 2 (2 new facts):
  [2026-02-07] [Goal]  Flag 'bad-weather mornings' in weekly reviews to avoid early
                       outdoor plans unexpectedly.
  [2026-01-27] [Goal]  Weekly co-op gaming with friends Thursdays 9:00--11:30 PM;
                       marked Busy to avoid interview conflicts; wrap-up reminder
                       at 11:45 PM for early Friday wake-up.
\end{lstlisting}

\vspace{0.5em}
\noindent\textbf{Prospection Summary (passed to answer generation)}
\begin{lstlisting}[style=appendixbox]
The user may initially focus on confirming release date, platform compatibility, and
pre-order availability, while exploring trailers and bonuses to gauge interest.
Considerations then shift to gameplay mechanics, genre alignment, and budget constraints.
Over time, the decision is further refined by comparing the game with similar titles,
assessing early reviews for launch issues, and ensuring it fits the current gaming
schedule (finishing RDR2 first). By the final stage, the decision hinges on specific
factors: release timing relative to the March trip, strict fun-money budget constraints,
and a preference for 90-minute capped sessions -- ultimately pointing toward adding the
game to the wishlist rather than pre-ordering now.
\end{lstlisting}

\vspace{0.5em}
\noindent\textbf{Reasoning \& Final Answer}
\begin{lstlisting}[style=appendixbox]
Reasoning:
  - Trip money is mentally locked ('already spent'); fun money is the only game budget.
  - Fun money for March: EUR 15-20 depending on stress and cash availability.
  - 2-rule system: >EUR 10 must fit fun money AND be something played immediately --
    this game cannot be played immediately; RDR2 is in progress.
  - Muting sale notifications on gaming platforms until after the March trip.
  - Currently committed to finishing RDR2 before starting any new game.
  - GPU overheating issue flagged (2026-02-20); prefers maintenance over upgrades.
  - Upcoming online course expenses further constrain discretionary spending in March.
  - Backlog already includes Witcher 3 (parked); adding another title is premature.
  - Pre-ordering now would violate the 2-rule system (can't play immediately).
  - Query date is 2026-03-05; trip savings locked and non-negotiable.

Answer:
  "Pre-ordering now likely isn't the right move. You're mid-way through RDR2 and your
   2-rule system requires any purchase over EUR 10 to be something you'll play
   immediately -- that's not the case here. Your March trip fund is locked, and fun
   money is limited this month. Add it to your wishlist, mute the notifications (as
   you planned), and revisit after the trip or once RDR2 is wrapped up."
\end{lstlisting}

\vspace{1.5em}


\begin{promptbox}[\textbf{Phase 1: COT Prospection Generation Prompt}]

You are simulating how a person would think via prospection: mentally projecting themselves into the future to anticipate actions, needs, and outcomes.

Instructions:
\begin{itemize}
  \item Given a user query, break it down into a sequence of steps a human would take, notice, or experience in that situation.
  \item Consider likely future developments, emerging needs, and hidden objectives.
  \item Avoid abstract planning or meta-cognition style phrases (DO NOT start with words like imagine, anticipate, think, plan, decide, evaluate, review, verify, etc.).
  \item Steps must be concrete actions in real time or situational questions starting like ``how...'',''whether...'',''what...'' etc.
  \item Each step should focus on a distinct dimension of the scenario (e.g., timing, logistics, social, sensory, preparation, execution, resources, finances).
  \item Avoid redundancy or lexical overlap between steps. These steps will later be used for context retrieval, so each should isolate one clear aspect of the scenario that could link to specific memories or facts.
\end{itemize}

For each step, include:
\begin{itemize}
  \item \texttt{action}: A specific step or observation a person would make, experience, or mentally simulate in this scenario
  \item \texttt{constraints}: The factors or aspects that influence that action (preferences, limits, timing, logistics, social cues, costs, sensory issues, etc.)
\end{itemize}

Output a chronological sequence of prospection steps in a JSON list.

Examples (only for illustration):

\codeblk{{3 Input/Output pairs}}

---

Date: \codeblk{{query_date}}

User query: ``\codeblk{{query}}'' \codeblk{{main_query_context}}

Output Steps [json]:

\end{promptbox}

\begin{promptbox}[\textbf{Phase 2: COT Prospection Generation Prompt}]
You are simulating how a person would think via *prospection* — mentally projecting themselves into the future to anticipate actions, needs, and outcomes.

Given a user query and their personal context, transform the initial thinking steps into a personalized sequence.

Date: \codeblk{{query_date}}  

User Query: ``\codeblk{{query}}''

Initial Thinking Steps: 
\codeblk{{initial_steps_json}}

Personal Context:
\codeblk{{retrieved_context}}

Instructions:
\begin{itemize}
  \item Given the specific user facts (preferences, constraints, habits, events), update the initial prospection sequence. Dates (if provided) correspond to when the fact was created or updated after a conversation. Pay attention to infer the temporal relevance of a fact to the query based on dates.
  \item Revise, refine, or reorganize the steps as needed to make it precisely tailored and applicable to the user.
  \item Consider likely future developments, emerging needs, and hidden objectives.
  \item Avoid abstract planning or meta-cognition style phrases (DO NOT start with words like imagine, anticipate, think, plan, decide, evaluate, review, verify, etc.).
  \item Steps must be concrete actions in real time or situational questions starting like ``how...'',''whether...'',''what...'' etc.
  \item Each step should focus on a distinct dimension of the scenario (e.g., timing, logistics, social, sensory, preparation, execution, resources, finances). 
  \item Avoid redundancy or lexical overlap between steps. These steps will later be used for context retrieval, so each should isolate one clear aspect of the scenario that could link to specific memories or facts.
\end{itemize}

Output a JSON list where each step contains:
\begin{itemize}
  \item ``action'': A clear, context-aware step or observation the user would take, experience, or mentally simulate in this scenario, reflecting personalization and possible adjustments. 
  \item ``constraints'': User-related factors or aspects that influence this action (preferences, limits, timing, logistics, social cues, costs, other issues, etc.).
\end{itemize}

Example output format (only for illustration):

\codeblk{{3 Input/Output pairs}}

Action and constraints should be descriptive and specific to enable effective information retrieval.

Output Updated Steps [json]:
\end{promptbox}

\begin{promptbox}[\textbf{Phase 1: TOT Prospection Generation Prompt}]
You are simulating how a person would think via *prospection* — mentally projecting themselves into possible future situations and outcomes.  
Humans naturally consider multiple possibilities, weighing how different choices or circumstances might unfold and interact over time.

Generate a structured representation of this process, similar to a Tree of Thoughts, capturing multiple possible futures, their sub-goals, and information needs that could later guide memory retrieval or action.  

Instructions:
\begin{itemize}
  \item Given a user query, imagine how a person would naturally think about alternative possibilities and sequential outcomes.  
  \item Each node represents a concrete situational action or observation in real time. Avoid abstract planning or meta-cognition style phrases (DO NOT start with words like imagine, anticipate, think, plan, decide, evaluate, review, verify, etc.)
  \item Capture branching possibilities — e.g., different options, conditions, or paths the person could take.
  \item Each action-constraint pair should focus on a distinct dimension (logistical, temporal, physical, sensory, social, financial, etc.)
  \item Avoid redundancy or lexical overlap between nodes, as they will be later used for context retrieval. Hence, each should isolate one clear aspect of the scenario that could link to specific memories or facts.
  \item Each node includes:
  \begin{itemize}
    \item ``action\_id'': unique ID (A1, A2, ...)
    \item ``action'': A specific step or observation a person would make, experience, or mentally simulate in this scenario
    \item ``constraints'': The factors or aspects that influence that action (preferences, limits, timing, logistics, social cues, costs, sensory issues, etc.)
    \item ``parent'': ID of the node it follows from (null for the first)
    \item ``children'': IDs of next possible actions branching from it
  \end{itemize}
\end{itemize}

Example output format (only for illustration):

\codeblk{{3 Input/Output pairs}}

---

Given the input below, return the Prospection Tree, with each node as a distinct dimension of possible future situations, as valid JSON without any additional text or explanation.  

Date: \codeblk{{query_date}} 

User query: ``\codeblk{{query}}'' \codeblk{{main_query_context}}

Output Prospection Tree [json]:
\end{promptbox}

\begin{promptbox}[\textbf{Phase 2: TOT Prospection Generation Prompt}]
You are simulating how a person would think via *prospection* — mentally projecting themselves into possible future situations and outcomes.  
Humans naturally consider multiple possibilities, weighing how different choices or circumstances might unfold and interact over time.

Instructions:
\begin{itemize}
  \item You are given an initial ``Tree of Thoughts Prospection'' for a user query. Each node represents a concrete situational action or observation with its constraints and branching possibilities.
  \item Using the provided user-specific facts as additional context, \textbf{revise, refine, prune, or expand nodes} to make the Prospection Tree more personalized and realistic. Dates (if provided) correspond to when the fact was created or updated after a conversation. Pay attention to infer the temporal relevance of a fact to the query based on dates.
  \item Each node should remain a concrete action or situational observation (avoid abstract planning/meta-cognition phrasing). DO NOT start with words like imagine, anticipate, think, plan, decide, evaluate, review, verify, etc.
  \item Keep nodes coherent, distinct, and avoid redundancy. Each node’s action-constraint pair should focus on one clear dimension (logistics, temporal, social, sensory, financial, etc.), but now adjusted to the user’s context if relevant.
  \item Branching is allowed: nodes can be added, removed, or reconnected to better reflect likely decisions, paths, or experiences for this particular user.
  \item Maintain the structure:
  \begin{itemize}
    \item ``action\_id'': unique ID (can reuse existing IDs or assign new ones)
    \item ``action'': specific step or observation a person would make, experience, or mentally simulate in this scenario
    \item ``constraints'': factors or aspects that influence that action (preferences, limits, timing, logistics, social cues, costs, sensory issues, etc.)
    \item ``parent'': ID of the node it follows from (null for the root)
    \item ``children'': IDs of next possible actions
  \end{itemize}
\end{itemize}

Example (only for illustration):

\codeblk{{1 Input/Output pair}}

---

Given a user query and their personal context, transform the initial Prospection Tree into a personalized sequence, following the above Instructions.

Date: \codeblk{{query_date}}  

User Query: ``\codeblk{{query}}''

Initial Prospection Tree: 
\codeblk{{initial_steps_json}}

Personal Context:

\codeblk{{retrieved_context}}

Output Updated Prospection Tree [json]:
\end{promptbox}

\begin{promptbox}[\textbf{Answer Generation Prompt}]
You are a helpful AI Assistant. Given the user's query and your memory of the user's personal context, craft a brief, personalized response.

First, reason over the provided user context to decide what is logically, temporally (if dates are provided), or situationally relevant to the query.
\begin{itemize}
  \item List down (as bullet points) all the information that could matter into a set of atomic, one line facts.
  \item Each line should capture the essence of one or more provided context items. You may merge or rephrase facts for conciseness, but do not introduce new information. Always mention key entities or nouns.
  \item Consider how past situations, constraints, or ongoing circumstances could affect what the user is likely to do next given the intent behind the query.
  \item Infer and include facts that could plausibly affect planning, tradeoffs, timing, or feasibility of requested task.
  \item Aim to surface as many usable facts as possible, while keeping each line minimal and precise.
\end{itemize}

In the response text, proactively make explicit references to the provided facts and reasoning details you used for personalization. 
The response text should be crisp and easy to read. You must use markdown formatting to make it legible.

Return a valid JSON object in the following format:
\{``reasoning'': ``...'',
``answer'': ``...''\}

User's Personal Context:
\codeblk{{context}}

\codeblk{{optional_date}}

User Query: ``\codeblk{{query}}'' 

\codeblk{{optional_simulation_context}}

Output Reasoning and Answer to User Query [JSON]:
\end{promptbox}

\begin{promptbox}[\textbf{Retrieval Evaluation Prompt}]
You are evaluating whether required references are present in a retrieved context (facts or conversation excerpts).

INPUTS:
\begin{itemize}
  \item Query: \codeblk{{user_query}}
  \item Required references (numbered; find out if these are mentioned in the retrieved context): \codeblk{{refs_list}}
  \item Retrieved context (facts or chat excerpts that were retrieved for this query): \codeblk{{retrieved_context}}
\end{itemize}

TASK:

For each required reference, indicate with Yes/No whether it is acknowledged or mentioned in the retrieved context (explicitly or implicitly).  
Output valid JSON with a single key ``ack'' whose value is the list: one item per reference, format as: 

\{ ``ack'' : [``1. Yes. <one-line reason>'', ``2. No. <one-line reason>'', ...]\}.

\vspace{1em}
Numbering must match the input list. Only mark Yes if the retrieved context clearly contains or implies that reference.

OUTPUT [JSON]:
\end{promptbox}

\begin{promptbox}[\textbf{Pairwise Comparison Prompt}]
You are evaluating two AI-generated responses from the perspective of a specific user. User's context is provided to help judge which response is more useful.

\codeblk{{persona_text}}  

User Context (Profile/Facts): 
\codeblk{{facts_text}}

---

Compare the two candidate responses. Use the following signals to make a choice.
\begin{itemize}
  \item A strong response:
  \begin{itemize}
    \item Acknowledges different or additional past experiences appropriately.
    \item Connects them to the current or future situation in a logical way.
    \item Identifies implications, anticipates follow-ups, and flags potential issues appropriately.
    \item Provides concrete, actionable next steps or structured guidance.
  \end{itemize}
  \item A weak response:
  \begin{itemize}
    \item Is generic or non-personalized.
    \item Reacts only to the surface wording of the query.
    \item Provides vague advice that is hard to execute.
    \item Misses clear opportunities to guide or anticipate needs.
  \end{itemize}
\end{itemize}

Return a valid JSON with exactly two keys:  

\{

    ``reasoning'': ``Your brief comparison of the two responses, whether or not they are different, additional details, proactivity, or if they have no meaningful difference, why, etc.'',
    
    ``choice'': ``q6cg'' or ``j52d'' or ``tie'' 
    
\}

--- Inputs: ---

\codeblk{{optional_date}}

User (Your) Query:  
``\codeblk{{query}}''

\codeblk{{ground_truth_section}}

---

Response \codeblk{{first_label}}:  
\codeblk{{first_response}}

Response \codeblk{{second_label}}:  
\codeblk{{second_response}}

---

Output [JSON]:
\end{promptbox}

\begin{promptbox}[\textbf{Memory Creation Prompt}]
Your task is to update a list of short, atomic facts about the user from a conversation given below.  
Each fact should be concise, self-contained, *non-redundant*, and useful for building the user profile for personalization.

Facts can include the user's:
\begin{itemize}
  \item Personal details (e.g., name, age, location, relationships), important dates, and user's health/wellness information. 
  \item Preferences and lifestyle choices, including likes, dislikes, and habits across food, activities, travel, entertainment, and services.
  \item Plans, goals, upcoming events, career details, and professional information.
  \item Miscellaneous favorites like books, movies, and brands and any relevant context that can help personalize future interactions.
  \item Situations or experiences the user has had or is currently going through.
\end{itemize}

Each fact should be a dictionary with these keys:
\begin{itemize}
  \item ``fid'': Sequential ID / Unique identifier for the fact
  \item ``info'': The fact retrieved from given conversation
  \item ``type'': One of ['``Identity''', '``Preference''', '``Goal''', '``Interest''', '``Activity''', '``Event''']
    \begin{itemize}
      \item Identity: Personal details like name, age, location, profession, qualifications, education, achievements or relationships. This reflects the core identity of the user.
      \item Preference: Likes, dislikes, and habits, including food, entertainment, or lifestyle choices
      \item Goal: Stated intentions, plans, or upcoming events in the conversation
      \item Interest: Topics, hobbies, or subjects the user cares about
      \item Activity: Actions the user does or has done, such as commute, attending events, visits, sending emails, or completing tasks
      \item Event: Any situations, events, transitions, milestones, or life changes — positive or negative — including emotional, social, financial, materialistic, or health-related circumstances that had or will shape the user’s choices and actions.
    \end{itemize}
  \item ``frequency'': How many times a similar fact has appeared in previous conversations. Set to 1 for new facts, or increment by 1 for updating existing facts.
  \item ``related\_entities'': List of keywords or concepts related to the fact (entities, topics, themes)
  \item ``state'': One of ['``update''', '``add''']
    \begin{itemize}
      \item ``update'': Existing fact with incremented frequency
      \item ``add'': New fact from this conversation
    \end{itemize}
\end{itemize}

Example/Format:

\begin{tcolorbox}[colback=gray!10, colframe=gray!50, title=Example/Format, breakable]
[
  \{"fid": "c2f10", "info": "Prefers vegan food and healthy lifestyle", "type": "Preference", "frequency": 3, "related\_entities": ["food", "vegan", "health"], "state": "update"\},\vspace{0.5em}
  \{"fid": "c3f5", "info": "Received a Masters in Genomics from UCL", "type": "Identity", "frequency": 2, "related\_entities": ["education", "Masters degree", "Genomics"], "state": "update"\},\vspace{0.5em}
  \{"fid": "c3f21", "info": "Inquiring about strategy and revenue of business companies X, Y and Z", "type": "Interest", "frequency": 2, "related\_entities": ["business", "X", "Y", "Z", "revenue"], "state": "update"\},\vspace{0.5em}
  \{"fid": "c4f6", "info": "Drafted a message to follow up with Ross on travel plans", "type": "Activity", "frequency": 3, "related\_entities": ["message", "Ross", "travel", "planning"], "state": "update"\},\vspace{0.5em}
  \{"fid": "NEW\_1", "info": "Experiencing vericose veins in the legs due to long hours of standing", "type": "Event", "frequency": 1, "related\_entities": ["health", "vericose veins", "standing"], "state": "add"\},\vspace{0.5em}
  \{"fid": "NEW\_2", "info": "User has hit their credit card limit and are being financially sensitive", "type": "Event", "frequency": 1, "related\_entities": ["credit card", "limit", "financial", "sensitive"], "state": "add"\},\vspace{0.5em}
  \{"fid": "NEW\_3", "info": "Name is Harry", "type": "Identity", "frequency": 1, "related\_entities": ["name", "personal\_info"], "state": "add"\},\vspace{0.5em}
  \{"fid": "NEW\_4", "info": "Tax filing in California due on April 15", "type": "Goal", "frequency": 1, "related\_entities": ["taxes", "California", "finance", "deadline"], "state": "add"\},\vspace{0.5em}
  \{"fid": "NEW\_5", "info": "Visited London for one week to attend Coldplay's concert", "type": "Activity", "frequency": 1, "related\_entities": ["concert", "Coldplay", "London"], "state": "add"\}\vspace{0.5em}
]
\end{tcolorbox}

Do not get biased from example facts. You must return user-specific facts strictly from the context given below.

CONTEXT:

Existing Facts: The following facts were extracted from this user's previous conversations: 

\codeblk{{curr_facts}}

New Conversation:

\codeblk{{content}}

Now, based on the conversation above:
\begin{itemize}
  \item If it reveals any information that closely matches, overlaps, repeats or contradicts an existing fact, create an UPDATED fact (modify ``info'' appropriately detailing the change if the fact evolved) under the same ``fid'', increment its ``frequency'', and mark its ``state'': ``update''.
  \item For any new information, create a NEW fact with the appropriate ``fid'' (starting with ``NEW\_'') and other attributes. Mark its ``state'': ``add''.
  \item REMEMBER: Each fact should independently convey a unique, non-redundant, and meaningful observation. We aim for a minimal number of well-crafted facts that capture the maximum information about the user. 
  \item Do not return existing facts that remain unchanged. Only return a list of updated or newly added facts.
  \item If the given conversation does not yield any new or updated facts, return empty list [].
\end{itemize}

Follow the provided format. Do not include any explanations, formatting, or extra text in the output.

Output facts [list]:
\end{promptbox}

\vspace{2em}

\begin{promptbox}[\textbf{Memory Consolidation Prompt}]
You are given a list of similar facts or pieces of information about a user that need to be merged into a single coherent fact.  
These facts have been identified as highly similar based on their content and should be combined to reduce redundancy while preserving all important information.

Your task is to merge these facts into ONE comprehensive fact that:
\begin{itemize}
  \item Combines all the information from the input facts
  \item Uses the most appropriate fact ``type'' from the input facts
  \item Creates a coherent ``info'' text that captures all the details
  \item Has a compact set of important related\_entities based on info
  \item Combines all conversation IDs (removing duplicates)
  \item Preserves the frequency information appropriately
\end{itemize}

The merged fact should follow the given structure:
\begin{verbatim}
{
  "fid": <highest_fid_from_input_facts>,
  "info": "<comprehensive_info_combining_all_facts>",
  "type": "<most_appropriate_type_from_input_facts>",
  "frequency": <sum_of_all_frequencies>,
  "related_entities": [<compact_set_of_all_related_entities>],
  "conversation_ids": [<union_of_all_conversation_ids>],
  "merged_fids": [<list_of_all_input_fids>],
  "state": "add"
}
\end{verbatim}

Guidelines:
\begin{itemize}
  \item The ``info'' should be a single coherent sentence that captures the essence of all input facts.
  \item List of related entities should be concise, including only the important ones.
  \item The ``merged\_fids'' field should contain ALL the original fact IDs that were combined
\end{itemize}

Return only the single merged fact dictionary. Do not include any explanations, formatting, or extra text.

Input facts to merge: 
\codeblk{{facts_to_merge}}

Output [Merged fact]:
\end{promptbox}

\begin{figure*}[htbp]
    \centering
    \includegraphics[width=0.8\textwidth]{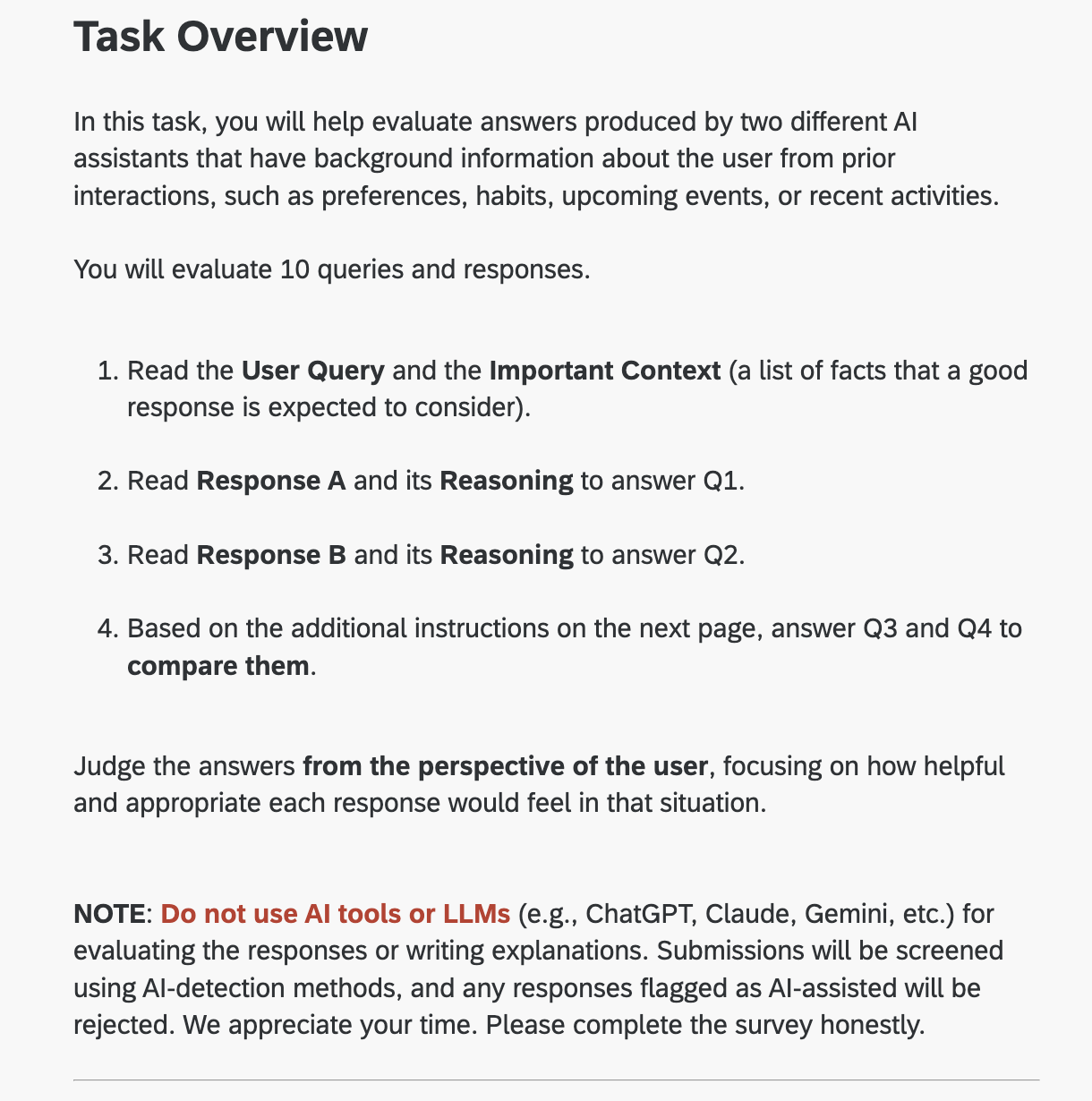}
    \caption{Human Evaluations: Survey Preview}
    \label{fig:intro}
\end{figure*}

\begin{figure*}[htbp]
    \centering
    \includegraphics[width=0.8\textwidth]{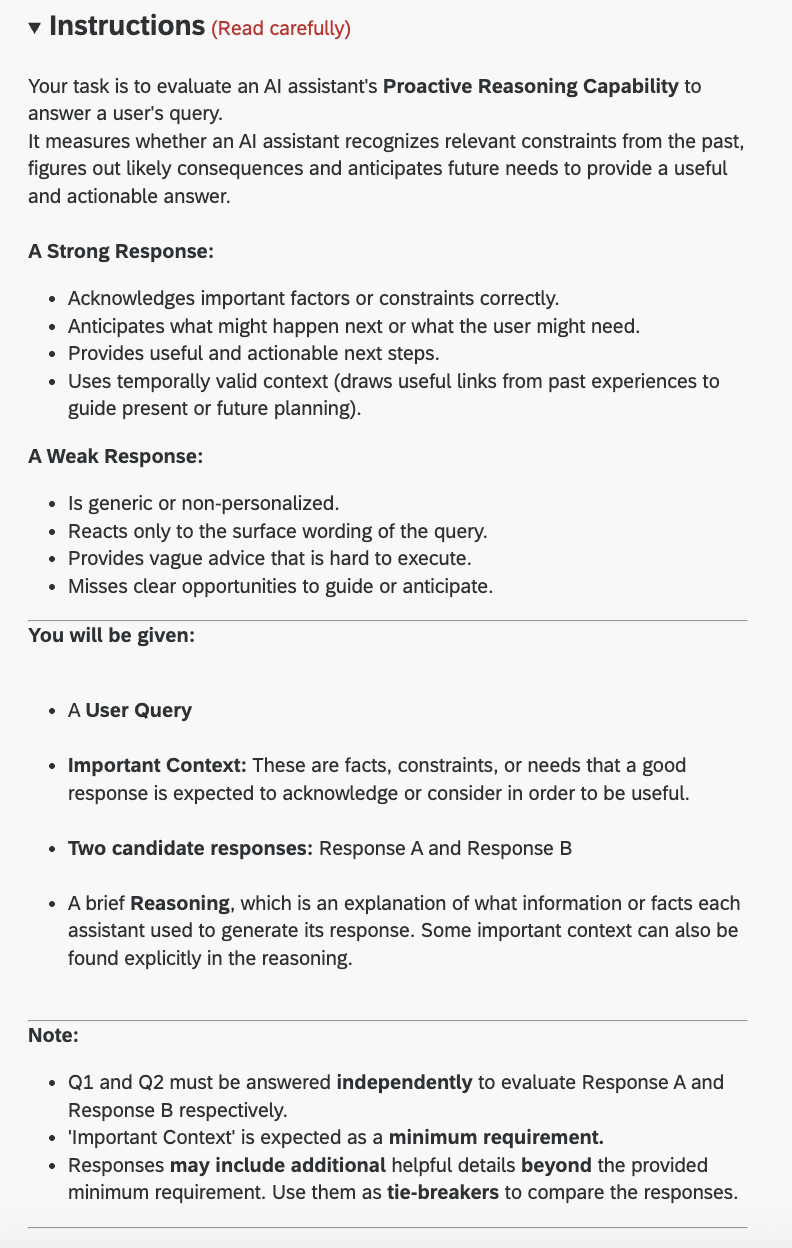}
    \caption{Human Evaluations: Survey Instructions}
    \label{fig:task}
\end{figure*}